\documentclass[aps,prd,twocolumn,superscriptaddress]{revtex4-1}

\usepackage{graphicx} 
\usepackage{color} 

\usepackage[bookmarks=true,bookmarksnumbered=true,colorlinks=true,linkcolor=blue,citecolor=blue,filecolor=blue,urlcolor=blue]{hyperref}  

\usepackage{amsfonts}
\usepackage{amsmath}
\usepackage{bm}
\newif\ifoneauthor
\oneauthortrue
\newcommand{\unit}[1]{\ {\rm #1}}

\newcommand{\Section}[1]{section \ref{#1}}
\newcommand{\Eq}[1]{Eq. (\ref{#1})}
\newcommand{\Table}[1]{Table \ref{#1}}
\newcommand{\Fig}[1]{Figure \ref{#1}}
\DeclareMathAlphabet{\mathpzc}{OT1}{pzc}{m}{it}
\usepackage{here} 
\definecolor{gray}{gray}{0.4}

\begin{document}

\title{Polarization test of gravitational waves from compact binary coalescences}


\author{Hiroki Takeda}
\email[]{takeda@granite.phys.s.u-tokyo.ac.jp}
\affiliation{Department of Physics, University of Tokyo, Bunkyo, Tokyo 113-0033, Japan}
\author{Atsushi Nishizawa}
\affiliation{Kobayashi-Maskawa Institute for the Origin of Particles and the Universe,Nagoya University, Nagoya, Aichi 464-8602, Japan}
\author{Yuta Michimura}
\affiliation{Department of Physics, University of Tokyo, Bunkyo, Tokyo 113-0033, Japan}
\author{Koji~Nagano}
\affiliation{KAGRA Observatory, Institute for Cosmic Ray Research, University of Tokyo, Kashiwa, Chiba, 277-8582, Japan}
\author{Kentaro Komori}
\affiliation{Department of Physics, University of Tokyo, Bunkyo, Tokyo 113-0033, Japan}
\author{Masaki Ando}
\affiliation{Department of Physics, University of Tokyo, Bunkyo, Tokyo 113-0033, Japan}
\author{Kazuhiro Hayama}
\affiliation{Department of Applied Physics, Fukuoka University, Nanakuma, Fukuoka 814-0180, Japan}




\date{\today}

\begin{abstract}
Gravitational waves have only two polarization modes in General Relativity. However, there are six possible modes of polarization in metric theory of gravity in general. The tests of gravitational waves polarization can be tools for pursuing the nature of space-time structure. The observations of gravitational waves with a world-wide network of interferometric detectors  such as Advanced LIGO, Advanced Virgo and KAGRA will make it possible to obtain the information of gravitational wave polarization from detector signals. We study the separability of the polarization modes for the inspiral gravitational waves from the compact binary coalescences systematically. Unlike other waveforms such as burst, the binary parameters need to be properly considered. We show that  the three polarization modes of the gravitational waves would be separable with the global network of three detectors to some extent, depending on signal-to-noise ratio and the duration of the signal. We also show that with four detectors the three polarization modes would be more easily distinguished by breaking a degeneracy of the polarization modes and even the four polarization modes would be separable. 
\end{abstract}


\maketitle

\section{Introduction}
 The first detection of the gravitational wave (GW) from binary black holes (BBH) by the Advanced LIGO (aLIGO) \cite{Aasi2015}  marked the dawn of the new field of gravitational-wave astronomy \cite{Abbott2016c}. After that, the observations of GWs by aLIGO and Advanced Virgo (AdV) \cite{Acernese2015} enabled some experimental studies to probe into the nature of gravity \cite{Abbott2016e,Abbott2017}. 

The accurate information of GW modes is expected to bring more knowledge about gravity. Polarizations of GWs can be  treated by Newman-Penrose formalism strictly and transparently \cite{Newman1962a, Alves2010a}. In general relativity (GR), a GW has two polarization modes (plus and cross modes called tensor mode). However a general metric theory allows the GW to have at most six polarizations: two tensor modes (plus, cross), two vector modes (vector x, vector y), and two scalar modes(breathing, longitudinal) \cite{Eardley1973a, Will1993}. In modified gravity theories such as the scalar-tensor theory \cite{Brans1961} and $f(R)$ gravity \cite{DeFelice2010}, scalar polarizations are allowed in addition to tensor modes. In contrast, at most six polarizations are allowed in bimetric gravity theory \cite{Visser1997} and at most five polarizations are allowed in massive gravity theory \cite{Rubakov2008}. If nontensorial polarization modes are found by the observations of GWs, it indicates that an alternatives to GR should exist as a fundamental theory and the theory of gravity should be extended beyond GR. Therefore, the test of gravitational waves polarizations can be a powerful tools for pursuing the nature of space-time structure.

In principle, the number of the detectors needs to be equal to the number of the polarization modes of the GW to separate the polarization modes.  In the near future, KAGRA \cite{Somiya2012, Aso2013, Akutsu2018, Akutsu2017}, a laser interferometric detector being developed at the Kamioka mine in Japan, and LIGO India will participate in aLIGO-AdV network as the fourth or fifth detector. Therefore, more polarization modes can be probed with the larger number of detectors.

Some analytical attempts to separate the polarization modes were made for the burst of the GW \cite{Hayama2013a}, the stochastic GW \cite{Nishizawa2009a} and the continuous waves of the GW \cite{Isi2015}. However, it is more difficult to analyze in the case of the compact binary coalescences (CBC) because the waveforms of the gravitational waves from CBC have the source model parameters, which determine the frequency evolution in time and are correlated each other. We focus on especially the polarization modes for the inspiral GW from the CBC. A simple data analysis method using the simple sine-Gaussian wavepacket waveform as a toy model was reported to probe gravitational wave polarization from CBC \cite{Isi2017a, Svidzinsky2017}.
It is necessary to gain a better understanding of the correlations and degeneracies among the parameters in a realistic waveform of CBC to separate and reconstruct the polarization modes in the presence of nontensorial polarization modes. We study parameter estimation errors and the correlations between parameters in the presence of nontensorial polarizations in addition to tensor modes with global detector networks systematically.

This paper is organized as follows. In \Section{analytical_method}, we describe polarization modes of gravitational waves, antenna pattern functions and detector signal. In \Section{inclination}, we explain the angular dependence of a gravitational  waveform in modified gravity and introduce  the polarization models adopted in our analysis. In \Section{setup}, we provide the basic of the Fisher analysis and introduce the our  numerical setup. In \Section{result}, we show the results of our parameter estimation in the presence of nontensorial polarization modes. We devote the last \Section{conclusion} to the conclusion of this paper.

\section{Antenna pattern functions}
\label{analytical_method}
\subsection{Polarization mode of  gravitational waves}
In general, there are six possible modes of polarization in a metric gravity theory  \cite{Eardley1973a, Will1993}. At any given space-time point, GW metric perturbations can be expressed as
\begin{equation}
\label{gw}
h_{ab}(t,\hat{\Omega})=h_{A}(t)e^{A}_{ab}(\hat{\Omega}),
\end{equation}
where $\hat{\Omega}$ is the sky direction of a GW source, $A=+,\times, x, y, b, l$ is polarization indices and are referred to plus, cross, vector x, vector y, breathing, and longitudinal, respectively. $e^{A}_{ab}(\hat{\Omega})$ are polarization tensors defined by
\begin{equation}
e^{+}_{ab}=\hat{e}_{x}\otimes\hat{e}_{x}-\hat{e}_{y}\otimes\hat{e}_{y},
\end{equation}
\begin{equation}
e^{\times}_{ab}=\hat{e}_{x}\otimes\hat{e}_{y}+\hat{e}_{y}\otimes\hat{e}_{x},
\end{equation}
\begin{equation}
e^{x}_{ab}=\hat{e}_{x}\otimes\hat{e}_{z}+\hat{e}_{z}\otimes\hat{e}_{x},
\end{equation}
\begin{equation}
e^{y}_{ab}=\hat{e}_{y}\otimes\hat{e}_{z}+\hat{e}_{z}\otimes\hat{e}_{y},
\end{equation}
\begin{equation}
e^{b}_{ab}=\hat{e}_{x}\otimes\hat{e}_{x}+\hat{e}_{y}\otimes\hat{e}_{y},
\end{equation}
\begin{equation}
e^{l}_{ab}=\sqrt{2}\hat{e}_{z}\otimes\hat{e}_{z},
\end{equation}
where the set of unit vectors $\{\hat{e}_x, \hat{e}_y, \hat{e}_z\}$ forms the wave orthonormal coordinate such that $\hat{e}_z=-\hat{\Omega}$ is a unit vector in the direction of propagation of the GW and $\hat{e}_z=\hat{e}_x\times \hat{e}_y$. We have a degree of freedom of choice $\hat{e}_x, \hat{e}_y$ around $\hat{e}_z$ axis. This degree of freedom is referred as the polarization angle $\psi_p$. 

\subsection{Antenna pattern functions and detector signal}
 The detector signal of the I-th detector is given by \cite{Tobar1999, Nishizawa2009a, Hayama2013a}
 \begin{equation}
 \label{detector_signal}
 h_I(t,\hat{\Omega})=d_{I}^{ab}h_{ab}(t)=F_I^{A}(\hat{\Omega})h_A(t).
 \end{equation} 
 Here $d_I$ is the detector tensor defined by 
  \begin{equation}
d_I:=\frac{1}{2}(\hat{u}_I\otimes\hat{u}_I-\hat{v}_I\otimes\hat{v}_I),  
  \end{equation}
  where $\hat{u}_I, \hat{v}_I$ are unit vectors along with arms of the I-th interferometric detector.
$ F_{I}^A$ is the antenna pattern functions of the I-th detector for polarization "A" defined by
  \begin{equation}
  \label{antenna}
  F_I^{A}(\hat{\Omega}):=d_I^{ab}e^{A}_{ab}(\hat{\Omega}).
  \end{equation}
  The general concrete formulas of the antenna pattern are provided in \cite{Nishizawa2009a}.
  We note that the above expressions are correct when the length of the interferometer arm is much smaller than the wavelength of observed GW. This condition is satisfied for ground-based detectors such as aLIGO, AdV, and KAGRA. 
 
\section{Polarizations}
\label{inclination}
\subsection{Angular dependence of a GW waveform in modified gravity}
The waveform of GW ($h_A(t)$ in \Eq{detector_signal}) depends on the inclination angle $\iota$ of a compact binary orbit as we show below. 
 In GR in which only  tensor modes are admitted,  the signal including the angular parameters (inclination angle, source position angles, detector position angles) and antenna pattern functions of I-th detector is
 \begin{equation}
 \label{tensor_signal}
 h_I=\frac{2}{5}\mathcal{G}_{T,I}h_{\rm{GR}},
 \end{equation}
 where $h_{\rm{GR}}$ corresponds to the GW waveform predicted by GR: $h_{\rm GR}$ is the Fourier component of the amplitude $h_{+}$ in \Eq{gw} in the direction perpendicular to the binary orbital plane. 
 Here $\mathcal{G}_{T,I}$ is the geometrical factor for the tensor mode for $I$-th detector, defined by
 \begin{eqnarray}
 \label{geo_tensor}
 \mathcal{G}_{T,I}:=\frac{5}{2}\{(1+\cos^2{\iota})F_{+,I}(\bm{\theta_s}, \bm{\theta_e})\nonumber \\ 	+2i\cos{\iota}F_{\times,I}(\bm{\theta_s},\bm{\theta_e})\}
 e^{i \phi_{D,I}(\theta_s,\phi_s,\theta_e,\phi_e)},
 \end{eqnarray}
 where  $\bm{\theta_s}:=(\theta_s,\phi_s,\psi_p)$ is source direction angle parameters $(\theta_s,\phi_s)$ and polarization angle $\psi_p$, $\bm{\theta_e}:=(\theta_e,\phi_e, \psi)$ is detector location and orientation angle parameters, and $\phi_{D,I}$ is the Doppler phase for $I$-th detector \cite{Creighton2011, Berti2005}. The factor of $2/5$ in \Eq{tensor_signal} appears so that the angular average of \Eq{geo_tensor} gives unity.
 
 The antenna pattern functions for nontensorial modes can be defined in \Eq{antenna}.  However, when we discuss the problem of the separation of polarization modes, we need to consider  the inclination angle dependence for scalar modes and vector modes, which is fixed by the geometry of the binary stellar system. Metric perturbations of GW can be given by the quadrupole formula
 \begin{equation}
 h_{ab}(t, \bm{x})=\frac{1}{r}\frac{2G}{c^4}\ddot{M}_{ab}(t-r/c),
 \end{equation} 
 where $M_{ab}$ is the moment of a mass distribution. $a, b$ run over $1, 2, 3$ that correspond to the source coordinate $\{x_1, x_2, x_3\}$ so that the binary circular motion is included in $x_1-x_2$ plane.
 In GR,  transverse-traceless projection for $M_{ab}$ can fix the gauge freedom \cite{Maggiore2007}. As a result, plus and cross tensor modes are only admitted. In modified gravity, the gauge symmetry is not held, leading to additional degrees of freedom for a GW. Therefore, nontensorial modes are obtained by keeping non-transverse-traceless modes. According to \Eq{gw}, each nontensorial polarization of GW propagating in the direction (along the coordinate 3) perpendicular to the orbital plane are
 \begin{equation}
 h_x(t)=\frac{1}{r}\frac{2G}{c^4}\ddot{M}_{13}(t-r/c),
 \end{equation}
 \begin{equation}
 h_y(t)=\frac{1}{r}\frac{2G}{c^4}\ddot{M}_{23}(t-r/c),
 \end{equation}
  \begin{equation}
 h_b(t)=\frac{1}{r}\frac{G}{c^4}(\ddot{M}_{11}(t-r/c)+\ddot{M}_{22}(t-r/c)),
 \end{equation}
 \begin{equation}
 h_l(t)=\frac{1}{r}\frac{\sqrt{2}G}{c^4}\ddot{M}_{33}(t-r/c).
 \end{equation}
For a GW propagating in the direction of $\hat{n}=(\sin{\iota}\cos{\phi},\sin{\iota}\sin{\phi},\cos{\iota})$, the expression of the amplitude can be obtained  by rotating $M_{ab}$ in the above expressions. 





For a circular binary stars moving as 
\begin{eqnarray}
x_1(t)&=&R\cos{(\omega_s t+\pi/2)},\nonumber\\
x_2(t)&=&R\sin{(\omega_st+\pi/2)},\\
x_3(t)&=&0, \nonumber
\end{eqnarray}
the second time derivatives of the mass moments are
\begin{equation}
\ddot{M}_{11}=-\ddot{M}_{22}=2\mu R^2\omega_{s}^2\cos{2\omega_st},
\end{equation}
\begin{equation}
\ddot{M}_{12}=2\mu R^2\omega_{s}^2\sin{2\omega_st},
\end{equation}
where $\omega_s$, $\mu$, $R$ are the  angular frequency of binary stars,  the reduced mass, and  the orbital radius, respectively.

Finally, we can get the simple expressions about the amplitudes for nontensorial polarization modes
\begin{equation}
h_x=-\frac{4G\mu\omega_{s}^2R^2}{rc^4}\frac{\sin{2\iota}}{2}\cos{(2\omega_{s}t_{\rm{ret}}+2\phi)},
\end{equation}
\begin{equation}
h_y=-\frac{4G\mu\omega_{s}^2R^2}{rc^4}\sin{\iota}\sin{(2\omega_{s}t_{\rm{ret}}+2\phi)},
\end{equation}
\begin{equation}
h_b=-\frac{4G\mu\omega_{s}^2R^2}{rc^4}\frac{\sin^2{\iota}}{2}\cos{(2\omega_{s}t_{\rm{ret}}+2\phi)},
\end{equation}
\begin{equation}
h_l=\frac{4G\mu\omega_{s}^2R^2}{rc^4}\frac{\sin^2{\iota}}{\sqrt{2}}\cos{(2\omega_{s}t_{\rm{ret}}+2\phi)},
\end{equation}
where $t_{\rm{ret}}$ is the retarded time defined by $t_{\rm{ret}}=t-r/c$. 

 From these expressions we define the geometrical factors for vector modes ($V_x,V_y$), scalar modes($S_2$), including the inclination angle dependence as follows.\\
 
 \begin{equation}
 \mathcal{G}_{V_x,I}:=\sqrt{\frac{525}{56}}\sin{2\iota}F_{V_x,I}(\bm{\theta_s},\bm{\theta_e})e^{i \phi_{D,I}(\theta_s,\phi_s,\theta_e,\phi_e)},
 \end{equation}
 \begin{equation}
  \mathcal{G}_{V_y,I}:=\sqrt{\frac{15}{2}}\sin{\iota}F_{V_y,I}(\bm{\theta_s},\bm{\theta_e})e^{i \phi_{D,I}(\theta_s,\phi_s,\theta_e,\phi_e)},
  \end{equation}
 \begin{equation}
  \mathcal{G}_{S_2,I}:=\sqrt{\frac{225}{8}}\sin^2{\iota}F_{b,I}(\bm{\theta_s},\bm{\theta_e})e^{i \phi_{D,I}(\theta_s,\phi_s,\theta_e,\phi_e)}.
  \end{equation}     
  
In addition to the above geometrical factors, the scalar dipole radiation may exist in modified gravity theories with a scalar degree of freedom. For a circular binary motion, the monopole radiation of the scalar mode vanishes and the dominant radiation of the scalar mode in the early inspiral phase can be dipole radiation.    The dependence on the inclination angle is proportional to $\sin{\iota}$, which is the result in the case of Brans-Dicke theory obtained by \cite{Chatziioannou2012}. Thereby we define the geometrical factor for the scalar dipole radiation as
\begin{equation}
    \mathcal{G}_{S_1,I}:=\sqrt{\frac{45}{2}}\sin{\iota}F_{b,I}(\bm{\theta_s},\bm{\theta_e})e^{i \phi_{D,I}(\theta_s,\phi_s,\theta_e,\phi_e)}.
\end{equation}  

These geometrical factor are normalized by angular average over the whole-sky and the inclination angle as well as \Eq{geo_tensor}.

\subsection{Polarization models}
\label{model}
Here we summarize polarization models used in our analysis. We assume the waveforms of nontensorial polarization modes are the same as those of tensor modes $h_{\rm GR}$ in \Eq{tensor_signal}, though these waveforms actually depend on the specific  theory of gravity. In other words, we consider pessimistic cases in terms  of separating polarization modes because it is more difficult to separate modes having the same waveforms.
 
\subparagraph{Model T: General Relativity model}
This model is a pure GR model. Thereby no additional polarization parameters is taken into account.
\begin{equation}
h_I=\mathcal{G}_{T,I}h_{\rm{GR}}.
\end{equation}

\subparagraph{Model TS1: tensor-scalar dipole model }
In this model, we add a scalar mode having the inclination-angle dependence as dipole radiation.
So an additional model parameter is the scalar mode amplitude $A_{S_1}$. 
\begin{equation}
h_I=\{\mathcal{G}_{T,I}+A_{S_1}\mathcal{G}_{S_1,I}\}h_{\rm{GR}}.
\end{equation}

\subparagraph{Model TS2: tensor-scalar quadrupole model }
In this model, we add a scalar mode having the inclination-angle dependence as quadrupole radiation. 
So an additional model parameter is the scalar mode amplitude $A_{S_2}$. 
\begin{equation}
h_I=\{\mathcal{G}_{T,I}+A_{S_2}\mathcal{G}_{S_2,I}\}h_{\rm{GR}}.
\end{equation}

\subparagraph{Model TVxS2: tensor-scalar quadrupole and vector x  model}
In this model, we add the combination of scalar mode having the inclination-angle dependence as quadrupole radiation and vector x mode. So additional model parameters are the amplitudes $(A_{S_2}, A_{V_x})$. 
\begin{equation}
h_I=\{\mathcal{G}_{T,I}+A_{S_2}\mathcal{G}_{S_2,I}+A_{V_x}\mathcal{G}_{V_x,I}\}h_{\rm{GR}}.
\end{equation}

\subparagraph{Model TVyS1: tensor-scalar dipole and vector y  model}
In this model, we add the combination of scalar mode having the inclination-angle dependence as dipole radiation and vector y mode. So additional model parameters are the amplitudes $(A_{S_1}, A_{V_y})$. It  is assumed that it is difficult to separate modes $S_2$ and $A_{V_x}$ than other combinations because these modes have the same inclination dependence of the geometrical factors. So we choose this combination as a pessimistic case. 
\begin{equation}
h_I=\{\mathcal{G}_{T,I}+A_{S_1}\mathcal{G}_{S_1,I}+A_{V_y}\mathcal{G}_{V_y,I}\}h_{\rm{GR}}.
\end{equation}

\subparagraph{Model TV: tensor vector model }
In this model, we add the combination of vector x and vector y mode. So additional model parameters are the amplitudes $(A_{V_x}, A_{V_y})$.
\begin{equation}
h_I=\{\mathcal{G}_{T,I}+A_{V_x}\mathcal{G}_{V_x,I}+A_{V_y}\mathcal{G}_{V_y,I}\}h_{\rm{GR}}.
\end{equation}

\section{Setup}
\label{setup}
\subsection{Fisher Analysis} 
  The model parameter estimation can be evaluated by a Fisher information matrix \cite{Finn1992, Cutler1994,Creighton2011}.
The Fisher information matrix $\Gamma$ is calculated by
\begin{equation}
\Gamma_{ij}:=4{\rm{Re}}\int^{\rm{f_{max}}}_{\rm{f_{min}}}df\sum_I \frac{1}{S_{n,I}(f)}\frac{\partial h^{*}_I(f)}{\partial\lambda^i}\frac{\partial h_I(f)}{\partial\lambda^j},
\end{equation} 
 where $S_{n,I}(f)$ is the I-th detector noise power spectrum and $\lambda_i$ is the i-th parameter. 
The root mean square of a parameter and the correlation coefficient between two parameters can be calculated using  the inverse of the Fisher information matrix.
 The root mean square of $\Delta\lambda^i$ is defined by
 \begin{equation}
(\Delta\lambda_i)_{\rm rms}:=\sqrt{\langle\Delta\lambda^i\Delta\lambda^i\rangle}=\sqrt{(\Gamma^{-1})^{ii}}, 
 \end{equation}
 and the correlation coefficient between $\lambda_i,\lambda_j$ is calculated by
 \begin{equation}
C(\lambda_i, \lambda_j):=\frac{\langle\Delta\lambda^i\Delta\lambda^j\rangle}{\langle(\Delta\lambda_i)^2\rangle\langle(\Delta\lambda_j)^2\rangle}=\frac{(\Gamma^{-1})^{ij}}{\sqrt{|(\Gamma^{-1})^{ii}(\Gamma^{-1})^{jj}|}},
 \end{equation}  
 where $\Delta\lambda^i$ is the measurement error of $\lambda^i$ and $\langle\cdot\rangle$ stands for ensemble average.
 
 The sky localization error is defined by
 \begin{equation}
 \Delta\Omega_s:=2\pi|\sin{\theta_s}|\sqrt{\langle(\Delta\theta_s)^2\rangle\langle(\Delta\phi_s)^2\rangle-\langle\Delta\theta_s\Delta\phi_s\rangle^2}.
 \end{equation}

Hereafter, we simply refer to $(\Delta\lambda_i)_{\rm rms}$ as $\Delta\lambda_i$, and call it the estimation error of $\lambda_i$.

\subsection{Analytical and Numerical setup}
We use the inspiral waveform up to 3 PN in amplitude and 3.5 PN in phase 
 \begin{equation}
 h_{\rm{GR}}=A_{\rm ins} e^{-i\phi_{\rm{ins}}},
 \end{equation}
 with
\begin{equation}
A_{\rm ins}=\frac{1}{\sqrt{6}\pi^{2/3}d_L}\mathcal{M}^{5/6}f^{-7/6}\sum_{i=0}^{6}(\pi\mathcal{M}f)^{i/3},
\end{equation}
\begin{equation}
\phi_{\rm{ins}}=2\pi ft_c-\phi_c-\frac{\pi}{4}+\frac{3}{128}(\pi\mathcal{M}f)^{-5/3}\sum_{i=0}^{7}\phi_i(\pi\mathcal{M}f)^{i/3},
\end{equation}
as a waveform of inspiral GW, compiled in \cite{Khan2016}.
Here $\mathcal{M}$ is the chirp mass, $d_L$ is the luminosity distance, $t_c$  is the coalescence time, and $\phi_c$ is the phase at the coalescence time.
We set the lower frequency end of various integration to be $f_{\rm{min}}=30 \unit{Hz}$ and the upper frequency end $f_{\rm{max}}$  to be  the frequency $f_{\rm{ISCO}}$ that is twice the innermost stable circular orbit frequency for a point mass in Schwarzschild spacetime
\begin{equation}
f_{\rm{ISCO}}=(6^{3/2}\pi M_{\rm tot})^{-1}\simeq0.0217M_{\rm{tot}}^{-1},
\end{equation}
where $M_{\rm{tot}}=m_1+m_2$ is the binary total mass.

 We consider 11 model parameters in GR
 \begin{equation}
(\log\mathcal{M},\log{\eta}, t_c, \phi_c, \log{d_L}, \chi_s, \chi_a, \theta_s, \phi_s, \cos{\iota}, \psi_p),
\end{equation}
 and additional  polarization amplitude parameters, for example  $A_{S_1}$ in the case of the model TS1. We assume that the fiducial values of the additional amplitude parameters are 1 in all models. Here $\log{\eta}, \chi_s, \chi_a$ are the logarithm of the mass ratio, the symmetric spin parameter and the antisymmetric parameter, respectively. We assume that the fiducial values of $t_c, \phi_c, \chi_s, \chi_a$ are 0 in all models.  We impose the priors for parameters having domain of definition; $\log{\eta}$, $\phi_c$, angular parameters $(\theta_s, \phi_s, \cos{\iota}, \psi_p)$ and spin parameters of binary compact stars $(\chi_s, \chi_a)$.
 
  We estimate model parameters for binary black holes (BBH) with equal mass $10M_{\odot}$ at $z=0.05$ and for binary neutron stars (BNS) with equal mass $1.4M_{\odot}$ at $z=0.01$ in each polarization model. Network total signal-to-noise ratio(SNR)$>8$ is required for all sources and angular parameters ($\cos{\theta_s},\phi_s,\cos{\iota},\psi_p$) are uniformly random. The number of sources calculated by fisher analysis for one model is 500.\\
  
  We consider two kinds of global networks composed of the two aLIGOs at Hanford and Livingston and AdV(HLV), and  HLV with KAGRA(HLVK).
aLIGOs and KAGRA are assumed to have their design sensitivity \cite{Abbott2016b}. AdV is also assumed to have its binary neutron star optimized sensitivity \cite{Abbott2016b}. \\

 We estimate model parameters in the case of BBH or BNS, with a detector network such as aLIGO-AdV(HLV) or aLIGO-AdV-KAGRA(HLVK) for each polarization model. 

\section{Results}
\label{result}

Our Fisher analysis results are shown in \Table{result_table}. We show the medians of parameter estimation errors of the luminosity distance, the sky localization, the additional polarization amplitude for each model.  We also show the medians of correlation coefficients larger than 10\% between the additional polarization amplitude and other parameters. In all models, the amplitude parameters of nontensorial polarization modes highly correlate with other amplitude parameters, $\ln{d_L}, \cos{\iota}$. 
\begin{table*}
\caption{Medians of parameter estimation errors and their correlation coefficients. Masses of BBH and BNS are $10M_{\odot}-10M_{\odot}$ and $1.4M_{\odot}-1.4M_{\odot}$ , respectively. Only correlation coefficients larger than $10\%$ are shown. The improvement factor is defined by the ratio of the error with HLV to the error with HLVK. We say that the polarization modes would be separable when the errors of the amplitudes parameter are less than unity. The two conditions for the separation of polarization modes are breaking the degeneracy among polarization modes by enough number of detectors and reducing the errors of the amplitude parameters from other practical point of view, for example signal-to-noise ratio(SNR) and the duration of the signal. }

  \begin{tabular}{|c|c||c|c|c||c|c|c|} \hline
   & parameter & BBH(HLV) & BBH(HLVK) &\shortstack{Improvement\\ Factor}& BNS(HLV) & BNS(HLVK) &\shortstack{Improvement\\ Factor}\\ \hline
		 &SNR 							& 33.3 	& 40.2 	&		& 36.4 	& 44.3	&\\ 
 ModelT 	&$\Delta\ln{d_L}$ 					& 0.269 	&  0.137	&1.96	& 0.183 	& 0.107	&1.71 \\ 
		&$\Delta\Omega_s[\rm{deg}^2]$ 		& 5.91 	&  1.77 	&3.34	& 1.39 	& 0.517	&2.69 \\ \hline      
		&$\Delta\ln{d_L}$ 					& 0.678 	&  0.179 	&3.79	& 0.359 	& 0.134	&2.68 \\ 
		&$\Delta\Omega_s[\rm{deg}^2]$ 		& 4.74 	&  0.912 	&5.20	& 0.919 	& 0.250	&3.68\\    
ModelTS1 &$\Delta A_{S1}$					& 1.16 	& 0.284 	&4.08	& 0.606 	& 0.197	&3.08\\ 
		&$C(A_{S1},\log{d_L})$				& 0.998 	&  0.989 	&		& 0.996 	&  0.984	&\\ 
		&$C(A_{S1},\cos{\iota})$				& -0.553 	& -0.500	&		&  -0.231 	& -0.159	&\\ \hline        
		&$\Delta\ln{d_L}$ 					& 0.676 	&  0.182 	&3.71	& 0.358 	& 0.134	&2.67 \\ 
    		&$\Delta\Omega_s[\rm{deg}^2]$ 		& 4.74 	&  0.913 	&5.09	& 0.862 	& 0.246	&3.50 \\    
ModelTS2 &$\Delta A_{S2}$					& 1.51 	& 0.385 	&3.92	& 0.765 	& 0.256	&2.99\\ 
		&$C(A_{S2},\log{d_L})$				&  0.997 	& 0.989  	&		& 0.996 	&  0.984	&\\ 
		&$C(A_{S2},\cos{\iota})$				& -0.609 	& -0.564 	&		&  -0.246 	& -0.189	& \\ \hline          
		&$\Delta\ln{d_L}$ 					& 1.58 	&  0.258 	&6.12	& 1.05 	& 0.190	&5.53 \\
    	   	&$\Delta\Omega_s[\rm{deg}^2]$ 		& 6.13 	&  0.885 	&6.92	& 0.783 	& 0.179	&4.37\\   
   		&$\Delta A_{S2}$					& 4.15 	& 0.486 	&8.54	& 2.48 	& 0.340	&7.29\\ 
ModelTVxS2 &$\Delta A_{V_x}$				& 2.23 	&  0.399 	&5.59	& 1.24 	& 0.228	&5.44\\ 
		&$C(A_{V_x},\log{d_L})$				& 0.945	&  0.690 	&		& 0.901 	& 0.633	& \\ 
  		&$C(A_{V_x},\cos{\iota})$				& 0.412 	& 0.360 	&		& -0.189	& -0.072	&\\ 
		&$C(A_{V_x},A_{S2})$				&  0.919 	& 0.576 	&		& 0.828 	& 0.557	&\\  \hline         
  		&$\Delta\ln{d_L}$ 					& 1.69 	&  0.253 	&6.68	& 1.05 	& 0.183	&5.74 \\
		&$\Delta\Omega_s[\rm{deg}^2]$ 		& 6.76 	&  0.879 	&7.69	& 0.831 	& 0.187	&4.44\\   
		&$\Delta A_{S1}$					& 3.72 	& 0.383 	&9.71	& 1.81 	& 0.273	&6.63\\ 
ModelTVyS1 &$\Delta A_{V_y}$				& 3.12 	&  0.389 	&8.02	& 1.75 	& 0.270	&6.48\\ 
		&$C(A_{V_y},\log{d_L})$				&  0.996 	& 0.990 	&		& 0.997 	& 0.986	&\\ 
		&$C(A_{V_y},\cos{\iota})$				& -0.660	& -0.322 	&		& -0.446 	& -0.010	&\\ 
		&$C(A_{V_y},A_{S1})$				&  0.996 	& 0.983 	&		& 0.996 	& 0.982	&\\  \hline    
		&$\Delta\ln{d_L}$ 					& 1.98 	&  0.310 	&6.39	& 1.22 	& 0.193	& 6.32\\ 
		&$\Delta\Omega_s[\rm{deg}^2]$ 		& 5.68 	&  0.795 	&7.14	& 0.813 	& 0.187	&4.35 \\  
		&$\Delta A_{V_x}$					& 2.55 	& 0.420 	&6.07	& 1.37 	& 0.241	&5.68\\ 
    ModelTV &$\Delta A_{V_y}$					& 3.91 	& 0.513 	&7.62	& 2.12 	& 0.298	&7.11\\ 
		&$C(A_{V_y},\log{d_L})$				&  0.999 	&  0.993 	&		& 0.998 	& 0.991	&\\ 
		&$C(A_{V_y},\cos\iota)$				&  -0.846	& -0.335 	&		& -0.307 	& -0.207	&\\ 
		&$C(A_{V_x},A_{V_y})$				&  0.987 	&  0.814 	&		& 0.948 	& 0.624	&\\ \hline
   
  \end{tabular}
\label{result_table}
\end{table*}

\begin{figure}
 \begin{center}
 \includegraphics[width=\hsize]{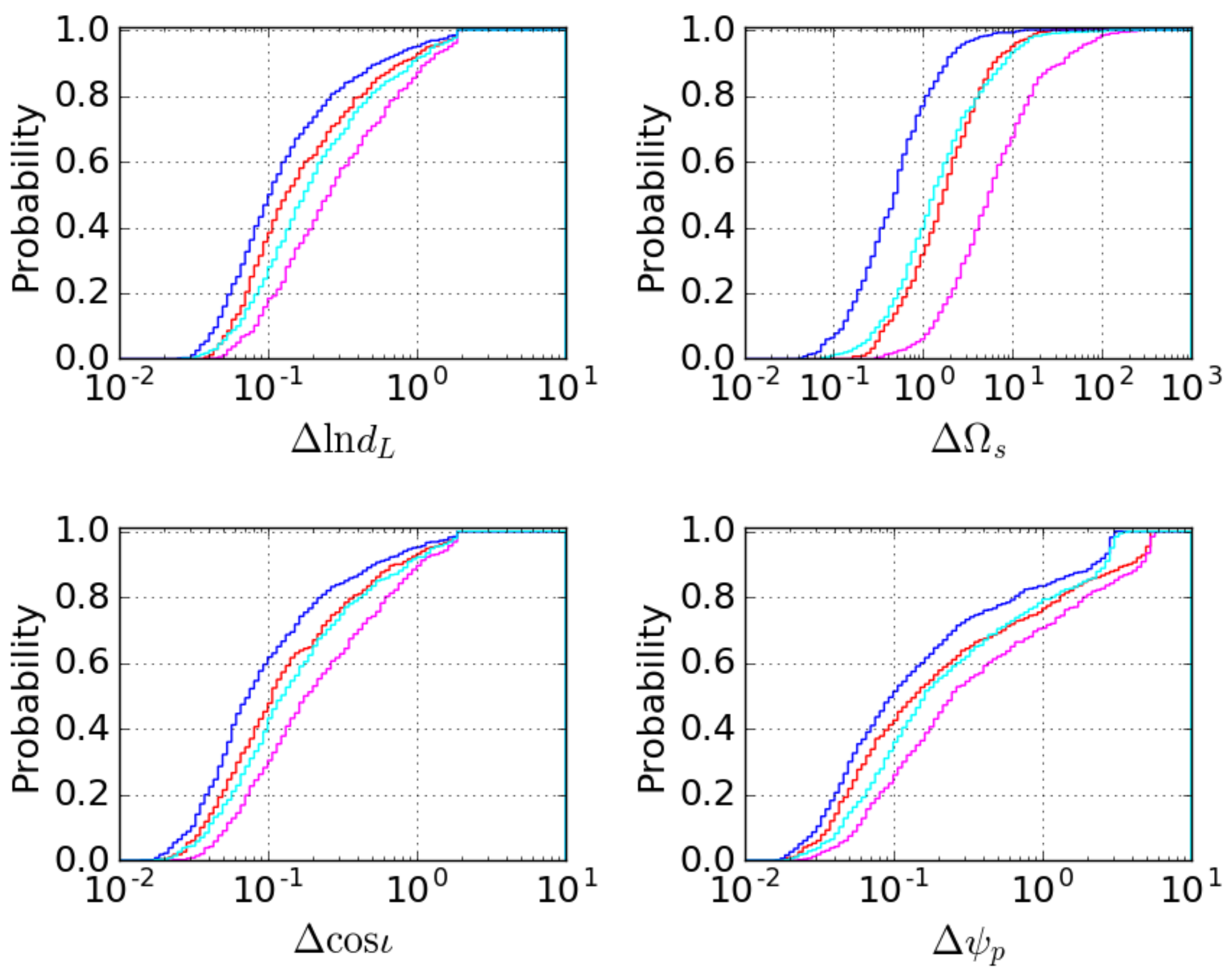}
 \end{center}
 \caption{Parameter estimation errors in the waveform model T.The colors are $10M_{\odot}-10M_{\odot}$ with HLV(magenta), $10M_{\odot}-10M_{\odot}$ with HLVK(red), $1.4M_{\odot}-1.4M_{\odot}$ with HLV(cyan), and $1.4M_{\odot}-1.4M_{\odot}$ with HLVK(blue). The reason for rapidly change of $\Delta\psi_p$ at around 3 is because we impose the priors for parameters having domain of definition, angular parameters and spin parameters of binary compact stars.}
 \label{modelT}
\end{figure}

The histograms of the parameter estimation errors in the Model T for the luminosity distance, the sky localization, the inclination angle, and the polarization angle are shown in \Fig{modelT}. 
In the model T, the errors  are improved by adding fourth detector, KAGRA. The error of the luminosity distance $\ln{d_L}$ is improved by about a factor of 2 and the sky localization error $\Omega_s$ is also improved by about a factor of 3. The other errors of the amplitude parameters, $\cos{\iota}, \psi_p$, are also improved by the fourth detector.

\begin{figure}
 \begin{center}
 \includegraphics[width=\hsize]{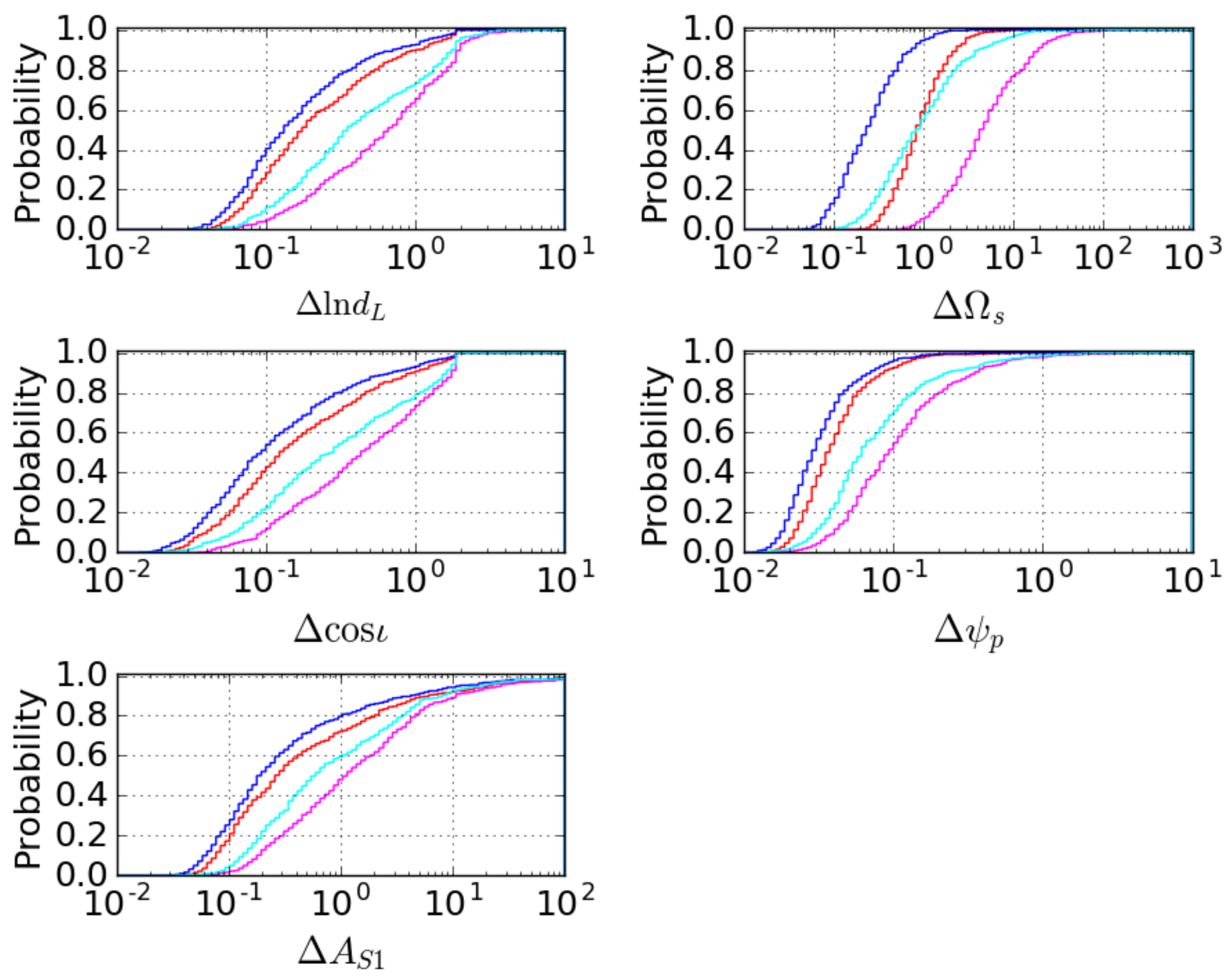}
 \end{center}
 \caption{Parameter estimation errors in the waveform model TS1.The colors are $10M_{\odot}-10M_{\odot}$ with HLV(magenta), $10M_{\odot}-10M_{\odot}$ with HLVK(red), $1.4M_{\odot}-1.4M_{\odot}$ with HLV(cyan), and $1.4M_{\odot}-1.4M_{\odot}$ with HLVK(blue).}
 \label{modelTS1}
\end{figure}

The histograms of the parameter estimation errors in the model TS1 for  the luminosity distance, the sky localization, the inclination angle, the polarization angle and the additional polarization amplitude are shown in \Fig{modelTS1}.
It is shown that the observation by the global network with 4 detectors can break a degeneracy among amplitude parameters. 
In the case of the model TS1, the errors of the amplitude parameters are more improved by adding the fourth detector, KAGRA than in the case of the model T.
For BNS, $A_s$ is determined even by three detectors, HLV. However, for BBH, it is difficult to separate the additional polarization mode. The error is much improved by adding the fourth detector, KAGRA. This suggests in the case of BBH that a four detectors global network is necessary to determine the mode though in principle three detectors can distinguish an additional scalar mode. 
The reason for the difference from the BNS case is because the signal of BBH is shorter than the signal of BNS. This results in the worse estimation error of the chirp mass, which is determined mainly from the phase of the signal. Indeed, $\Delta\ln{\mathcal{M}}=0.0019$ (median) in the case of BBH, but  $\Delta\ln{\mathcal{M}}=0.00015$ (median) in the case of BNS with HLV. Since the chirp mass is also included in the GW amplitude, it results in the worse parameter estimation of the amplitude in the case of BBH. On the other hand, $\Delta\ln{\mathcal{M}}=0.0017$ (median) in the case of BBH, but  $\Delta\ln{\mathcal{M}}=0.00014$ (median) in the case of BNS with HLVK. $\Delta\ln{\mathcal{M}}$ for BBH is improved by $11 \%$ while $\Delta\ln{\mathcal{M}}$ for BNS is improved by $7 \%$.

In the case of the model TS2, the errors and correlations behave the same way as in the case of the model TS1.

\begin{figure}
 \begin{center}
 \includegraphics[width=\hsize]{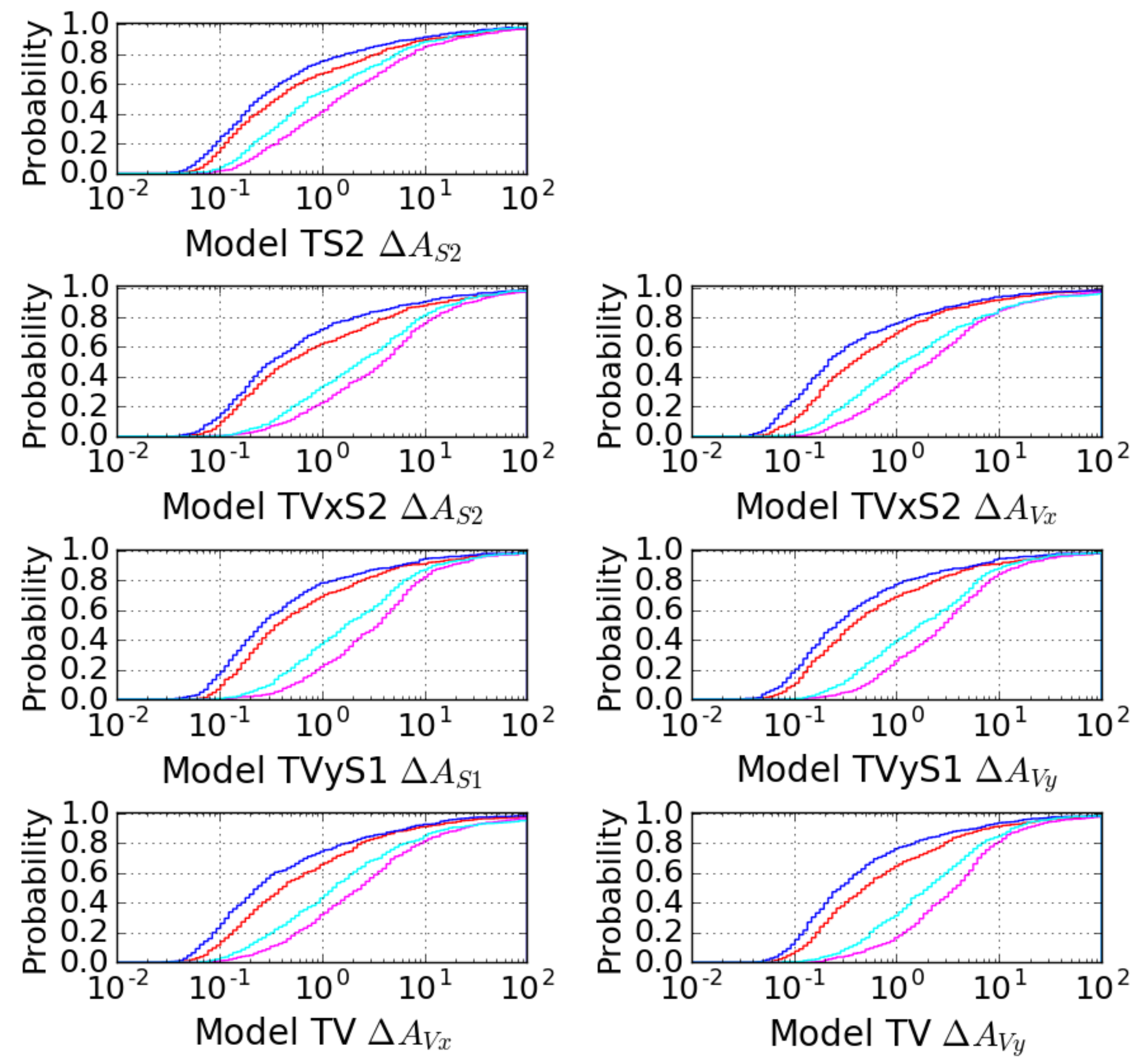}
 \end{center}
 \caption{Parameter estimation errors of the additional polarization amplitudes assumed as 1 for all models.The colors are $10M_{\odot}-10M_{\odot}$ with HLV(magenta), $10M_{\odot}-10M_{\odot}$ with HLVK(red), $1.4M_{\odot}-1.4M_{\odot}$ with HLV(cyan), and $1.4M_{\odot}-1.4M_{\odot}$ with HLVK(blue).}
 \label{pol_amp}
\end{figure}

Our results of parameter estimation for the additional  polarization amplitudes in other polarization models are shown in \Fig{pol_amp}.
In the model TVxS2, TVyS1 and TV, the errors of the amplitude parameters are larger than 1 with HLV for both BBH and BNS as shown in \Table{result_table}, so that  four detectors are always necessary to determine two additional polarizations. In these cases, the errors of the additional polarization modes are more improved by adding the fourth detector, KAGRA than in the case of the model TS1 and the model TS2. In all of these models, the error  is  more than 5 times improved for both BBH and BNS.

 In our analysis, the coalescence time and the phase at the coalescence time for the nontensorial mode are assumed to be those of the tensor mode. If these parameters of the nontensorial mode are introduced to the polarization models, it may affect the estimation errors. 
 We have checked how the modification of the coalescence time and the phase at coalescence affect the results by (1) changing the fiducial values of those parameters and by (2) introducing another set of those parameter for a nontensorial mode. However both (1) and (2) did not affect the final result of parameter estimation much. On the other hand, we also changed all the detector sensitivity by a factor of 10. The errors of the amplitude parameters in the model TS1 and TS2 with HLV were improved, but the errors in the model TVsS2, TVyS1 and TV were not improved with HLV because of the degeneracy among polarization modes. These indicate the polarization degrees of freedom are characterized by overall amplitude parameters and require the same number of detectors to separate the modes and extract the polarization information from the detector signal of GW in principle. 
 
\begin{figure}
 \begin{center}
 \includegraphics[width=\hsize]{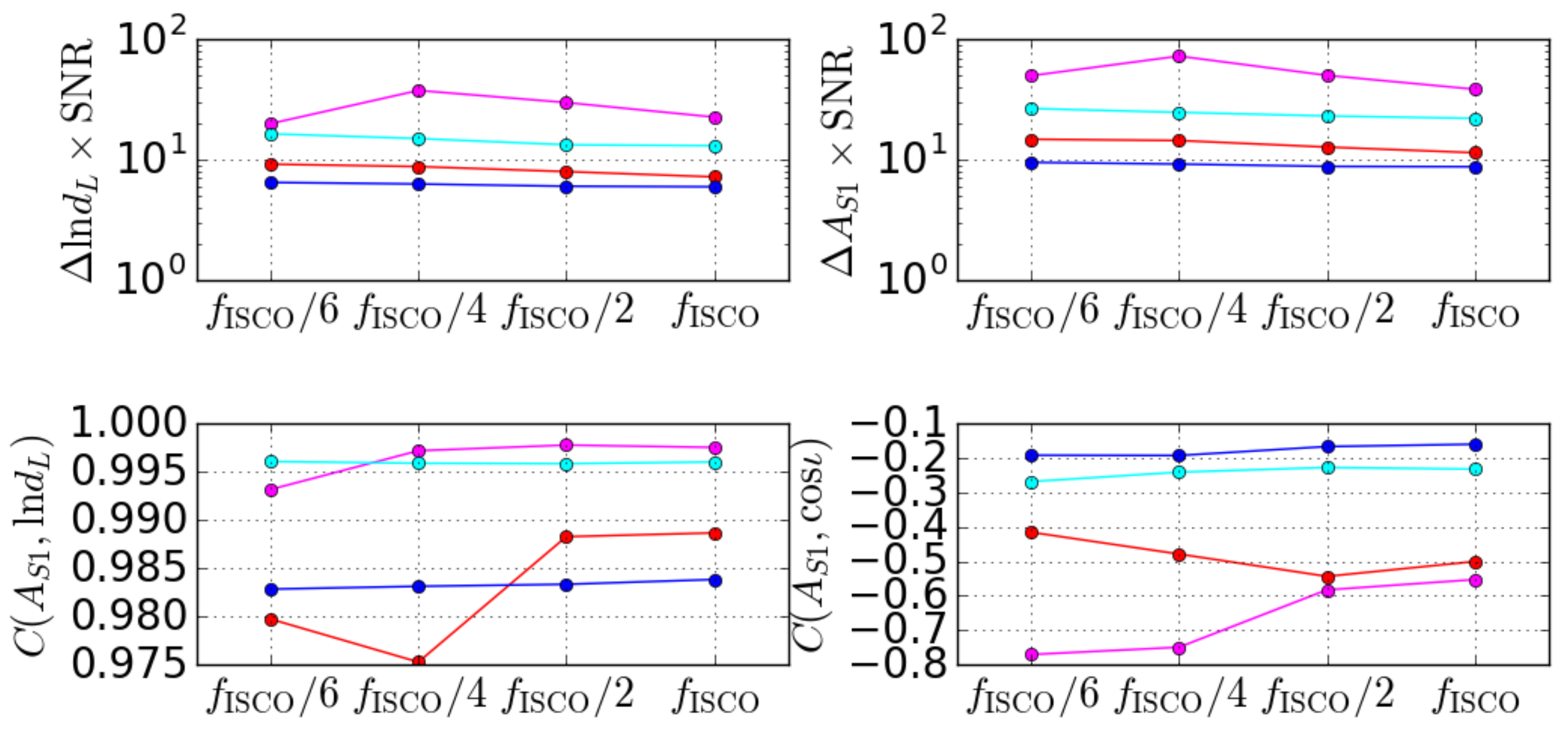}
 \end{center}
 \caption{The vertical axes are the estimation errors multiplied by SNR and the correlation coefficients in the case of the model TS1, and the horizontal one the upper frequency end used in evaluating the parameter estimation by Fisher analysis. The colors are $10M_{\odot}-10M_{\odot}$ with HLV(magenta), $10M_{\odot}-10M_{\odot}$ with HLVK(red), $1.4M_{\odot}-1.4M_{\odot}$ with HLV(cyan), and $1.4M_{\odot}-1.4M_{\odot}$ with HLVK(blue). }
 \label{fmax}
\end{figure}

\Fig{fmax} shows the maximum-frequency  dependence of the errors in the model TS1. We change the $f_{\rm max}$ to $f_{\rm ISCO}/2, f_{\rm ISCO}/4, f_{\rm ISCO}/6$ and plot the errors multiplied by SNR and correlation coefficients. For most cases, a change of $f_{\rm max}$ does not affect the amplitude estimation corrected by SNR because the errors of the scalar-mode amplitude multiplied by SNR are almost flat. This indicates that the change of $f_{\rm max}$ simply scales the amplitude error as well as SNR. The points dissatisfying the scaling, especially correlations in the case of $f_{\rm ISCO}/6\simeq35\unit{Hz},\ f_{\rm ISCO}/4\simeq52\unit{Hz}$ for BBH, would appear due to the short integration range compared to $f_{\rm min}=30\unit{Hz}$.

\begin{figure}
 \begin{center}
 \includegraphics[width=\hsize]{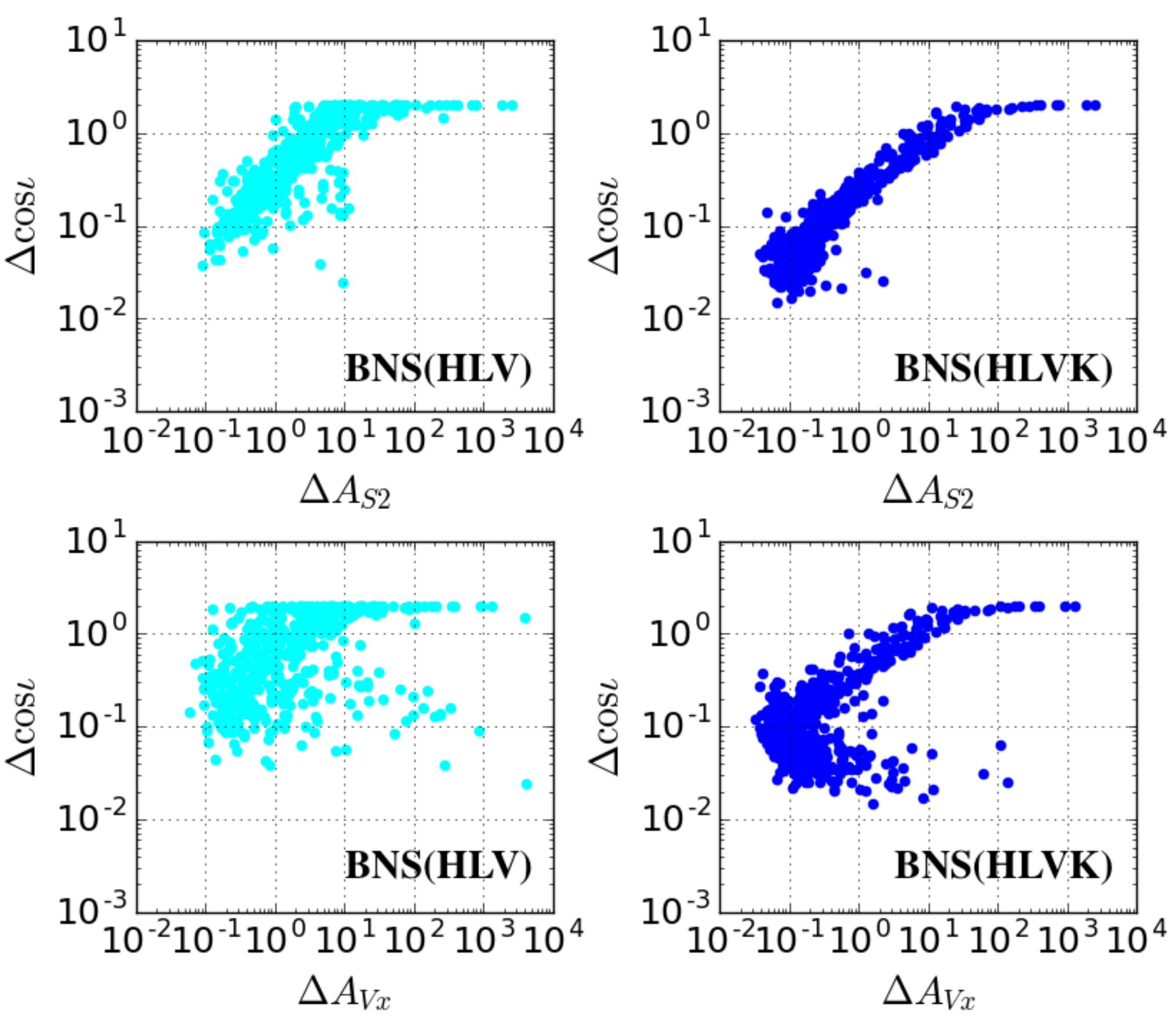}
 \end{center}
 \caption{Scatter plots of the error of the nontensorial polarization amplitudes vs the estimation error of the inclination angle for BNS in the model TVxS2. The colors are $1.4M_{\odot}-1.4M_{\odot}$ with HLV(cyan), and $1.4M_{\odot}-1.4M_{\odot}$ with HLVK(blue).  }
 \label{scatter2a}
\end{figure}

\begin{figure}
 \begin{center}
 \includegraphics[width=\hsize]{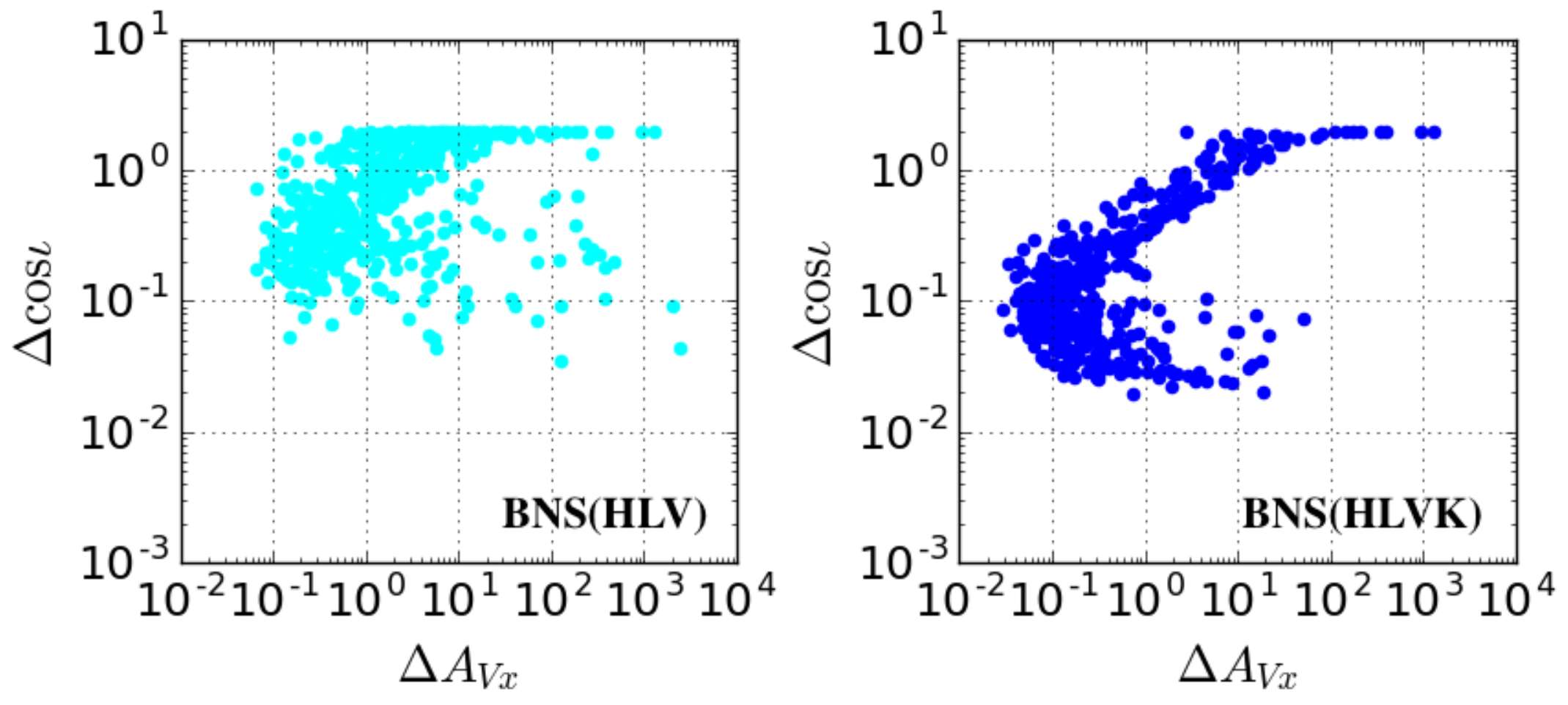}
 \end{center}
 \caption{Scatter plots of the error of the nontensorial polarization amplitude $A_{V_x}$ vs the estimation error of the inclination angle for BNS in the model TV. The colors are $1.4M_{\odot}-1.4M_{\odot}$ with HLV(cyan), and $1.4M_{\odot}-1.4M_{\odot}$ with HLVK(blue).  }
 \label{scatter3}
\end{figure}

\begin{figure}
 \begin{center}
 \includegraphics[width=\hsize]{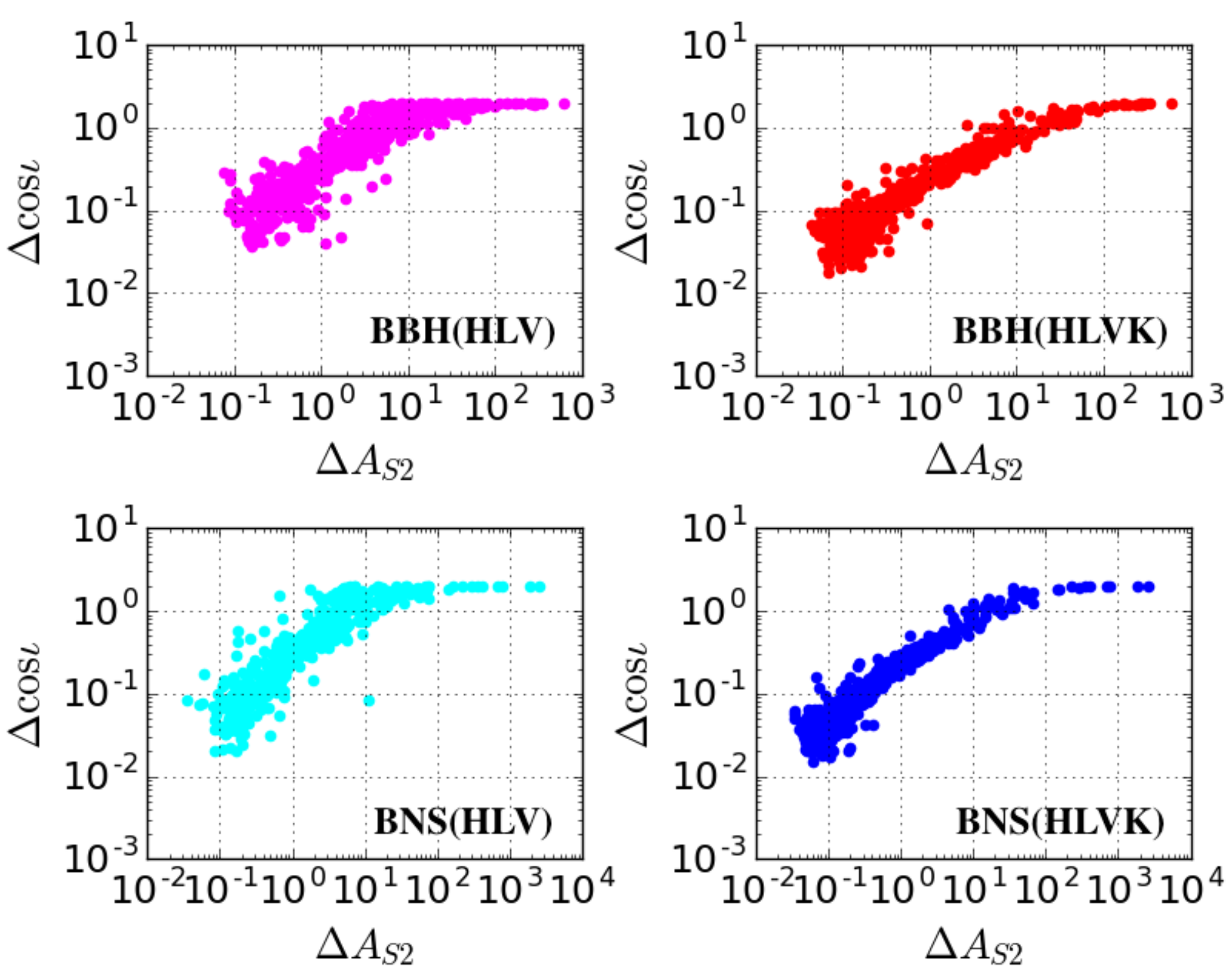}
 \end{center}
 \caption{A scatter plot of the error of the nontensorial polarization amplitudes vs the estimation error of the inclination angle in the model TS2. The colors are $10M_{\odot}-10M_{\odot}$ with HLV(magenta), $10M_{\odot}-10M_{\odot}$ with HLVK(red), $1.4M_{\odot}-1.4M_{\odot}$ with HLV(cyan), and $1.4M_{\odot}-1.4M_{\odot}$ with HLVK(blue).  }
 \label{scatter1b}
\end{figure}

The additional polarization amplitudes correlate with the inclination angle strongly as shown in \Table{result_table}. \Fig{scatter2a} is scatter plots of the errors of  the nontensorial mode amplitudes vs the error of the inclination angle for BNS in the model TVxS2. It is shown that the different polarization modes depend on the inclination angle differently. \Fig{scatter3} is a scatter plot $A_{V_x}$ vs the inclination angle for BNS in the model TV. The plots of $A_{V_x}$ in \Fig{scatter2a} and \Fig{scatter3} show that the same polarization mode depends on the inclination angle the same way even in different models. We have checked that $A_{V_y}$ in both the model TVyS1 and TV behaves the same way and the scalar modes $A_{S_1}, A_{S_2}$ also behave the same way in the different models. 
\Fig{scatter1b} is the scatter plots of the errors of the nontensorial polarization amplitude $A_{S_2}$ vs the estimation error of the inclination angle in the model TS2 for BBH and BNS. In both cases of BBH and BNS, the distributions of the plots have the same appearance. 

\section{Detection Limit}
\begin{figure}
 \begin{center}
 \includegraphics[width=\hsize]{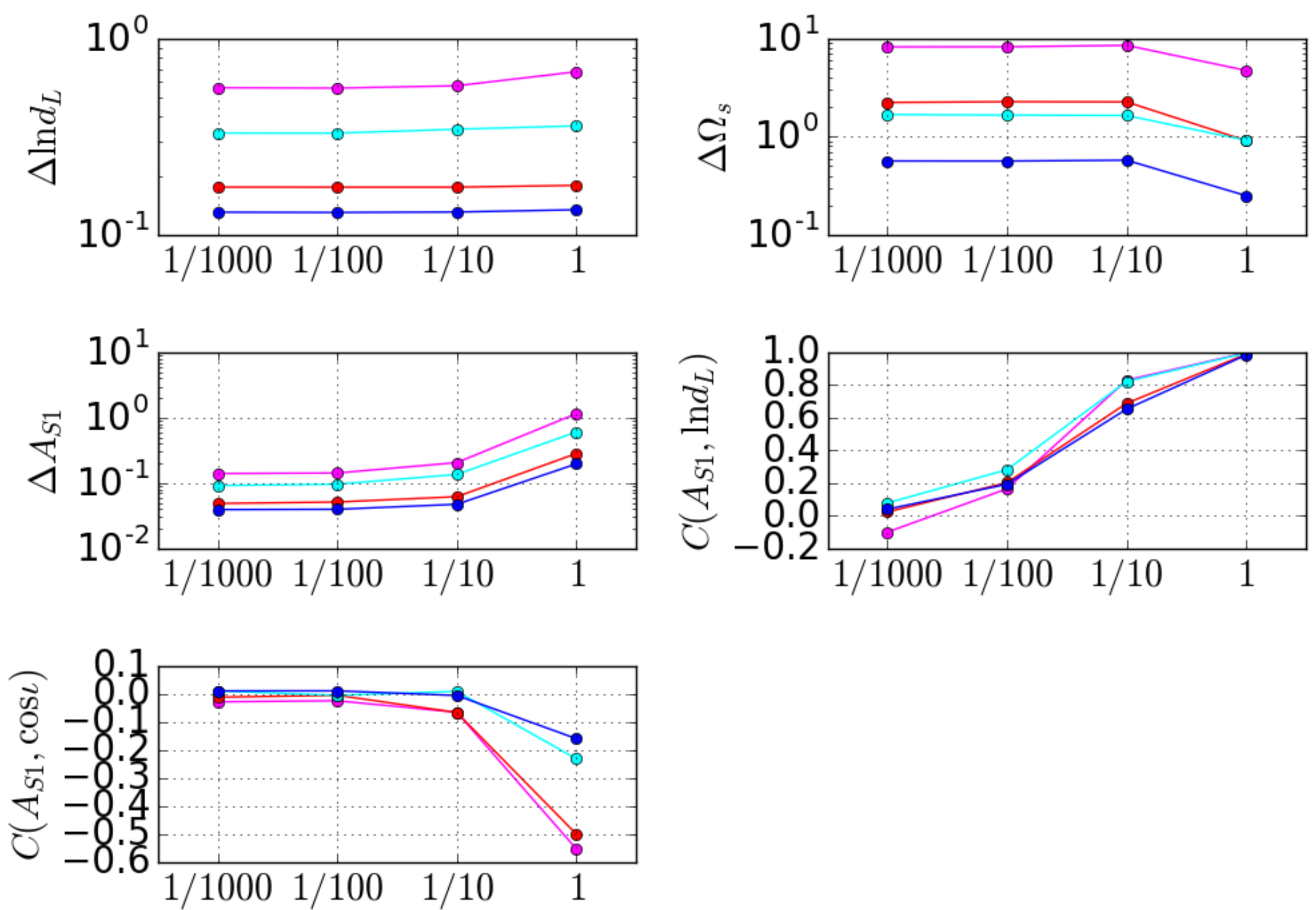}
 \end{center}
 \caption{The vertical axes are the estimation errors and the correlation coefficients in the case of the model TS1, and the horizontal one the fiducial values of the amplitude parameter $A_{S_1}$ used in evaluating the parameter estimation by Fisher analysis. The colors are $10M_{\odot}-10M_{\odot}$ with HLV(magenta), $10M_{\odot}-10M_{\odot}$ with HLVK(red), $1.4M_{\odot}-1.4M_{\odot}$ with HLV(cyan), and $1.4M_{\odot}-1.4M_{\odot}$ with HLVK(blue). }
 \label{change_fid_ts1}
\end{figure}

We assume that the fiducial values of all the additional amplitude parameters for nontensorial modes are unity in our analysis above because we first need to understand the correlations between model parameters to reconstruct the polarization modes from the detector signal. We changed the fiducial values to $1/1000, 1/100 ,1/10$ to show how the choice of the fiducial values affect the estimation errors. \Fig{change_fid_ts1} is the fiducial value  dependence of the errors and correlation coefficients in the model TS1. Since the error of the luminosity distance and the sky localization error are hardly changed, it indicates that the errors are determined by tensor modes mainly. Also the $A_{S_1}$ error is hardly  changed at the lower fiducial values than $1/10$. It implies that the detection limit of the $A_{S_1}$ is given roughly by $1/{\rm SNR}$. The correlation coefficients $C(A_{S_1}, {\rm ln}d_L), C(A_{S_1}, \cos{\iota})$ are smaller for the smaller  fiducial values. This also implies that it is difficult to detect the $A_{S_1}$ under the detection limit.

We also checked how the fiducial values affect the estimation errors in the other models TS2, TVxS2, TVyS1 and TV. The behavior of the estimation errors and the correlation coefficients are same as in model TS1. Under detection limit, which is given roughly by $1/{\rm SNR}$, the errors of the additional amplitude parameters are hardly changed and it is difficult to detect those parameters.

\section{Conclusion}
\label{conclusion}
We estimated model parameters of the gravitational waves from compact binary coalescences with detector networks such as aLIGO-AdV(HLV) or aLIGO-AdV-KAGRA(HLVK) for various polarization models in which the polarization degrees of freedom are characterized by overall amplitude parameters. We found that in principle at least the same number of detectors is required to separate the modes and extract the polarization information from the detector signal of gravitational waves. However, even if the number of detectors is equal to the number of the polarization modes, it is difficult to separate the modes in some cases, depending on the correlation among the amplitude parameters. Thereby there are two conditions for the separation of polarization modes; (i) the same number of detectors as the number of polarization modes and (ii) significant SNR and the long duration of the signal. In general, there is a strong correlation between the additional polarization amplitude and the inclination angle of the binary orbit. For the same polarization modes, the appearance and strength of the correlation between the additional polarization amplitude and the inclination angle are the same even in different models as long as a degeneracy among the amplitude parameters are broken.
 
The participation of the fourth detector in the network of the gravitational wave detectors will make it possible to extract the polarization information from detector signal of the gravitational waves generated by the compact binary coalescences even in the case of the presence of two nontensorial polarizations in addition to tensor modes. In some cases including only one nontensorial polarizations in addition to tensor modes, the separation of polarization modes is possible with the fourth detector by breaking a parameter degeneracy.

\section*{Acknowledgements}
H. T. and K. K. acknowledge financial support received from the Advanced Leading Graduate Course for Photon Science (ALPS) program at the University of Tokyo. H.T. is also supported by JSPS KAKENHI Grant
No. 18J21016. A. N. was supported by JSPS KAKENHI Grant No. JP17H06358. K.N. is supported by JSPS KAKENHI Grant  No. 17J01176. This work was supported by JSPS Grant-in-Aid for Scientific Research (B) No.  18H01224. This study was supported by MEXT, JSPS Leading-edge Research Infrastructure Program,  Grant-in-Aid for Scientific Research on Innovative areas (No. 2905, No. 17H06357, No. 17H06365). This work was supported by MEXT, JSPS Leading-edge Research Infrastructure Program, JSPS Grant-in-Aid for Specially Promoted Research 26000005, MEXT Grant-in-Aid for Scientific Research on Innovative Areas 24103005, JSPS Core-to-Core Program, A. Advanced Research Networks, the joint research program of the Institute for Cosmic Ray Research, University of Tokyo, National Research Foundation(NRF) and Computing Infrastructure Project of KISTI-GSDC in Korea, the LIGO project, and the Virgo project.


\bibliography{paper_pol_estimation}
\bibliographystyle{h-physrev3}
\end{document}